\documentclass[runningheads]{llncs}
\usepackage{todonotes}
\usepackage{graphicx}
\usepackage[caption=false]{subfig}
\usepackage{pdflscape}
\usepackage[british]{babel}
\usepackage{lscape}
\usepackage{tabularx} 

\usepackage{multirow}
\usepackage{booktabs}

\usepackage{array}
\newcolumntype{L}[1]{>{\raggedright\let\newline\\\arraybackslash\hspace{0pt}}m{#1}}
\newcolumntype{C}[1]{>{\centering\let\newline\\\arraybackslash\hspace{0pt}}m{#1}}
\newcolumntype{R}[1]{>{\raggedleft\let\newline\\\arraybackslash\hspace{0pt}}m{#1}}

\usepackage{hyperref}
\hypersetup{ colorlinks=false,       
    linkcolor=blue,         
    citecolor=blue,     
    filecolor=blue,    
    urlcolor=blue     
}

\begin{document}
\title{Technical Debt and Waste in Non-Functional Requirements Documentation: An Exploratory Study}
\titlerunning{Technical Debt and Waste in Non-Functional Requirements Documentation}
%
\author{Gabriela Robiolo\inst{1} \and
Ezequiel Scott\inst{2} \and
Santiago Matalonga\inst{3} \and
Michael Felderer\inst{4}
}
\authorrunning{Robiolo et al.}
%
\institute{LIDTUA (CIC), Facultad de Ingeniería, Universidad Austral, Argentina \\ \email{grobiolo@austral.edu.ar} \and
Institute of Computer Science, Tartu Unviersity, Estonia \\
\email{ezequiel.scott@ut.ee}\and
School of Computing, Engineering and Physical Sciences, University of the West of Scotland, United Kingdom \\
\email{santiago.matalonga@uws.ac.uk} \and
Department of Computer Science, University of Innsbruck, Austria\\
\email{michael.felderer@uibk.ac.at}}
\maketitle              
\begin{abstract}
\textit{Background:} To adequately attend to non-functional requirements (NFRs), they must be documented; otherwise, developers would not know about their existence. However, the documentation of NFRs may be subject to Technical Debt and Waste, as any other software artefact.
\textit{Aims:} The goal is to explore indicators of potential Technical Debt and Waste in NFRs documentation.
\textit{Method:} Based on a subset of data acquired from the most recent NaPiRE (Naming the Pain in Requirements Engineering) survey, we calculate, for a standard set of NFR types, how often respondents state they document a specific type of NFR when they also state that it is important. This allows us to quantify the occurrence of potential Technical Debt and Waste. 
\textit{Results:} Based on 398 survey responses, four NFR types (Maintainability, Reliability, Usability, and Performance) are labelled as important but they are not documented by more than 22\% of the respondents. We interpret that these NFR types have a higher risk of Technical Debt than other NFR types.
Regarding Waste, 15\% of the respondents state they document NFRs related to Security and they do not consider it important.
\textit{Conclusions:} There is a clear indication that there is a risk of Technical Debt for a fixed set of NFRs since there is a lack of documentation of important NFRs. The potential risk of incurring Waste is also present but to a lesser extent.
\end{abstract}
\section{Introduction}\label{introduction}

Non-functional requirements (NFRs) are of high importance for the success of a software project \cite{Eckhardt2016}. Nevertheless, there exists evidence that NFRs tend to come second class to functional requirements \cite{Eckhardt2016} \cite{Chung_leite_2009}. We see this as a pervasive problem, regardless of the methodology that the development process follows. Quality management models and standards like ISO 9001:2015 \cite{ISO90012015} and CMMI \cite{Chrissis2007} require that functional and non-functional requirements are documented as way of conveying their importance. These software development models and standards take a``do as you say, say as you do" approach where documentation and upfront planning is used to mitigate the risk of not delivering the software product within the constraints of the project. In fact, the ISO/IEC/IEEE 29148:2018 standard for software and systems requirements engineering \cite{ISOIECIEEE_29148}, prescribes that both functional and non-functional requirements have to be documented.

Agile software engineering highlights the need of ``continuous attention to technical excellence" \cite{agile_Manifest}. Agile software engineering methods mainly rely on immediate feedback and postulate the sufficient availability of knowledgeable software developers to mitigate potential quality risks. Unfortunately, agile values and principles often seem to be adopted na\"{i}vely \cite{Hoda2017}, i.e., equating agile with avoiding documentation \cite{Stettina2011}. 

The starting point of the research presented in this paper is the assumption that in order to be able to adequately handle NFRs, they must be documented --otherwise developers would not know about their precise nature or even their existence. Based on a subset of data acquired from the most recent NaPiRE (\textbf{Na}ming the \textbf{P}ain \textbf{i}n \textbf{R}equirenments \textbf{E}ngineering) survey conducted in 2018 \cite{Fernandez2018}, we calculate for the NFR types Compatibility, Maintainability, Performance, Portability, Reliability, Safety, Security, and Usability how often respondents state they document a specific type of NFR when they also state that this type of NFR is important. We address the following research questions:

\textbf{RQ1:} \textit{Can we identify Technical Debt and Waste in requirements documentation from the responses in the NaPiRE questionnaire?} To understand the current status of Technical Debt and Waste in the context of NaPiRE, we calculate the occurrence of potential Technical Debt (i.e., NFR not documented although labeled as important) and the occurrence of potential Waste (i.e., NFR documented although labeled as not important) with regard to the different NFR types (i.e. Compatibility, Maintainability, Performance, Portability, Reliability, Safety, Security, and Usability).

\textbf{RQ2: }\textit{How does the practitioners' context influence the occurrence of Technical Debt and Waste in requirements documentation?} The system type, the project size, and the type of development process are usually the first variables to be considered in exploratory studies. We explored the practitioners' responses related to these variables in order to understand how they influence the occurrence of Technical Debt and Waste in the context of NaPiRE.

Our results show a clear indication of Technical Debt in several NFRs, with Maintainability, Reliability, Usability, and Performance being the NFRs with the highest frequency of occurrence of potential Technical Debt. Furthermore, when breaking down the analysis by the type of development process, the development processes at the extremes of the spectrum (i.e., purely plan-driven or purely agile) alter the indication of the Technical Debt pattern. 
Furthermore, our results show that there is less risk of incurring in Waste. Less than 15\% of the respondents stated that they document NFR which they do not consider important. With Security being the NFR with the highest frequency of respondents at about 15\%.


\section{Background} \label{sec:background}

In this section, we introduce the \textbf{Na}ming the \textbf{P}ain \textbf{i}n \textbf{R}equirenments \textbf{E}ngineering (NaPiRE) initiative\footnote{NaPiRE web site -- http://napire.org} and present an overview on research about NFRs.

\subsection{The NaPiRE Project} \label{subsec:naire}
The objective of the NaPiRE project is to establish a comprehensive theory of requirements engineering (RE) practice and to provide empirical evidence to practitioners that helps them address the challenges of requirements engineering in their projects. These objectives shall be achieved by collecting empirical data in surveys conducted world-wide and in repeating cycles.
At the time of writing this paper, three rounds of the NaPiRE survey have been carried out. The first survey round was conducted in Germany and the Netherlands in 2012 \cite{MendezFernandez2013_NAPIREFIRST}. The second round, conducted in the years 2014 and 2015, was extended to ten countries \cite{Fernandez2017_NAPIREMAIN}. The third round of the survey was conducted in 2018 and collected data from 42 countries.
The research presented in this paper is based on the data collected in the third round. Since the NaPiRE survey instrument has evolved since 2014/2015, a direct comparison between past analysis results and currently ongoing analyses is not always possible. This holds especially for the topic of non-functional requirements covered in this paper, which is the first one published on data from the third run.
 
The previous installments of the NaPiRe survey have been successful in sparking complementing research into several viewpoint of requirements engineering. For instance, to compare requirements engineering practices across geographical regions \cite{Fernandez2015}\cite{kalinowski2016preventing} or by development method \cite{Wagner2017_CIBSE}\cite{wagner2018agile}. We argue that, although NaPiRE data has been extensively analyzed, it has so far not been analyzed with regards to practitioners' perceptions about handling NFRs.

\subsection{Published Research on Non-Functional Requirements} \label{Section: Related papers}

This section presents an overview of past directions in non-functional requirements (NFR) research with a focus on survey research in the context of software industry.

Borg et al. \cite{Borg2003} presented a case study on how NFRs are dealt with in two software development organisations. The authors interviewed 14 software developers in two organisations. Their results show that, in both contexts, functional requirements take precedence over non-functional requirements. 

Berntsson Svensson et al. \cite{Svensson2009} investigated the challenges for managing NFRs in embedded systems. They interviewed ten practitioners from five software companies. Their results show a widespread variation in how the respondents dealt with NFR, they also suggest a relationship between a lack of documentation of NFR and a dismissal of NFR during the project lifecycle.
Behutiye et al. \cite{Behutiye2017} investigated how software development teams using agile projects deal with non-functional requirements. The authors interviewed practitioners in four companies developing software with agile methodologies. Each company followed a different practices when documenting NFR (including not documenting them and relying on tacit knowledge). 

Ameller et al. \cite{Ameller2012} looked at how software architects deal with non-functional requirements. 
Their results highlight a lack of common vocabulary among software architects to convey NFR, the two most important NFR types were performance and usability, and that NFRs are often not documented, and when documented, the documentation was usually imprecise and was rarely maintained. Also, Proot et al. \cite{Poort_2012} presented a survey about the perceived importance of non-functional requirements among software architects. 
Their results suggest that architects consider NFRs important to the success of their software projects. 

De la Vara et al. \cite{delaVara:2011} present a questionnaire-based survey capturing the more important NFRs from the point of view of practitioners. 31 practitioners from 25 organizations were selected within the industrial collaboration network of the authors. The top five NFRs identified are Usability, Maintainability, Performance, Reliability, and Flexibility. 
Haigh et al. \cite{Haigh2010} empirically examined the requirements for software quality held by different groups involved in the development process. She conducted a survey of more than 300 current and recently graduated students of one of the leading Executive MBA programs in the United States, asking them to rate the importance of each of 13 widely-cited attributes related to software quality. The results showed the following ranking of NFRs: Accuracy, Correctness, Robustness, Usability, Integrity,  Maintainability,  Interoperability,  Augmentability, Efficiency, Testability,  Flexibility, Portability, Reusability. 

In summary, we conclude that the topic of NFRs has been extensively researched but there is few evidence of how the NFRs are documented. Furthermore, the specific relationship between importance level and degree of documentation has not yet been investigated. 

\section{Research Method} \label{sec:research_method}

Before describing the research method we first present our understanding of relevant concepts and assumptions about our research. Secondly, we introduce the terminology used in this paper. Then, we present our research questions. Finally, we describe the data extraction and analysis procedure that we followed in order to answer those research questions.

\subsection{Concepts and Assumptions} \label{Section: Concepts and assumption}
According to the software product quality standard ISO 25010:2011~\cite{ISOStandard25010_2011}, a non-functional requirement is a ``requirement that specifies criteria that can be used to judge the operation of a software system''~\cite{ISOStandard25010_2011}. The same standard defines a model for the evaluation of quality in use and product quality of a software system. Within this quality model, the product quality attributes (also known as quality characteristics) are defined. A quality attribute is a specification of the stakeholders' needs (Functional Suitability, Performance Efficiency, Compatibility, Usability, Reliability, Security, Maintainability, Portability). We argue that both terms, NFR and product quality attribute, are related and often used interchangeably in industry, even though this is not correct according to the precise definitions of these terms. Upon careful consideration, in particular looking at how the NaPiRE survey instrument framed the questions related to NFRs, in this work we interpret NFRs to be all requirements that do not specify a functional behaviour. Furthermore, we do not differentiate between NFR and product quality attribute. We claim that (1) the NaPiRE questionnaire has not made this distinction evident, (2) most practitioners would not care for the subtleties of this differentiation, and (3) interchangeable use of terms is pervasive among practitioners and researchers \cite{Behutiye2017}, \cite{Poort_2012}, \cite{Chung_leite_2009}. 
In order to be consistent with the survey instrument used in the NaPiRE survey, in this paper, we use the term ``quality attribute" instead of ``NFR" when we present our research questions and the results of our analyses. 

This research is driven by our assumption that, in agreement with \cite{ISOIECIEEE_29148}, both functional and non-functional requirements have to be documented. In the context of software quality assurance, which is defined in ANSI/IEEE Standard 729-1983 \cite{ANSI/IEEE7291983}, the confidence of the established technical requirements is achieved by checking the software and the documentation and verifying their consistency.

Therefore, the ideal situation is that when a quality attribute is considered as important for the development project, then it must be documented. To better convey this understanding we refer to the Technical Debt metaphor. Technical Debt, as defined by \cite{Seaman2011}, is ``a metaphor for immature, incomplete, or inadequate artefacts in the software development lifecycle that cause higher costs and lower quality in the long run. These artefacts remaining in a system affect subsequent development and maintenance activities, and so can be seen as a type of debt that the system developers owe the system." Also, Zengyang et al. \cite{Zengyang2015} pointed out that documentation of Technical Debt refers to insufficient, incomplete, or outdated documentation in any aspect of software development. That is, when practitioners perceive a quality attribute as important but fail to document requirements associated to the quality attribute, we will interpret this as an indication of the incurred Technical Debt. We differentiate from \cite{Ernst2012}, which defines Technical Debt in requirements as the distance between the implementation and the actual state of the world. 

We follow similar reasoning on the other end of the spectrum but we rely on the concept of Waste in Lean development. In Lean development, Waste is defined as anything that does not add value \cite{Ikonen_Waste}. In the domain of software development, the types of Waste can be interpreted as: extra features, waiting, task switching, extra processes, partially done work, movement, defects, or unused employee creativity \cite{Xiaofeng2012}. Therefore, when practitioners are investing effort in documenting requirements for artifacts (quality attributes) that they do not consider important, we are interpreting that such an effort could be better placed elsewhere in the development process, and understand it as a source of Waste.

In the most recent round of the NaPiRE survey, practitioners were asked about their perception of importance regarding a set of pre-defined quality attributes in the context of the project they were currently working on. In addition, they were asked whether they document quality attributes. The specific questions related to these aspects and their possible responses are shown in Table \ref{tab:items}. Question Q1 asks for the level of importance of each NFR type and Q2 for its degree of documentation. Questions Q3, Q4, and Q5 request the context factors project size, system type, and development process type, respectively. By combining the answers to Q1 and Q2, we can investigate if practitioners are following the requirements documentation recommendation for quality attributes in a specific context determined by Q3, Q4, and Q5.

\begin{table*}[]
\centering
\caption{NaPiRE Questionnaire items used for the analysis}
\label{tab:items}
\renewcommand{\arraystretch}{1}
\begin{tabular}{p{.05\textwidth}p{.45\textwidth}p{.33\textwidth}R{.15\textwidth}}
\toprule
\bfseries ID & \bfseries Questionnaire item & \bfseries Possible responses & \bfseries Variables                    \\ \midrule
Q1 & Are there quality attributes which are of particularly high importance for your development project? If yes, which one(s)? & Compatibility, Maintainability, Performance, Portability, Reliability, Safety, Security, Usability                        & v\_6-v\_13                   \\ \midrule
Q2 & Which classes of non-functional requirements do you explicitly consider in your requirements documentation? & Compatibility, Maintainability, Performance, Portability, Reliability, Safety, Security, Usability                                     & v\_97-v\_102, v\_303, v\_103 \\ \midrule
Q3 & How many people are involved in your project?                                                                              & Free text                                                                                                           & v\_3                         \\ \midrule
Q4 & Please select the class of systems or services you work on in the context of your project.                                 & Software-intensive embedded systems (SIES), Business information systems (BIS), Hybrid of both software-intensive embedded systems and business information systems (HYB) & v\_4                         \\ \midrule
Q5 & How would you personally characterize your way of working in your project?                                                 & Agile, Rather agile, Hybrid, Rather plan-driven, Plan-driven                                                                                           & v\_24                        \\ \bottomrule
\end{tabular}
\end{table*}

Table \ref{Table:Important_Documented} conveys our perception of the possible scenarios. In the ideal world, practitioners do not incur in Technical Debt (Important and Not Documented), nor do they Waste effort in documenting requirements which they do not consider important (Not Important and Not Documented (NI\_ND)). However, our experience leads us to expect that, practitioners are restricted by the context of their development projects and they are bound to incur in Technical Debt and Waste. In this research, we will look for evidence of this understanding in the responses to the NaPiRE 2018 survey.

\begin{table}[h]
\renewcommand{\arraystretch}{1.2}
\centering
\caption{Perception of importance and availability of documentation quadrant.}
\label{Table:Important_Documented}
\begin{tabular}{|p{.2\columnwidth}|p{.39\columnwidth}|p{.39\columnwidth}|}
\hline
                                      & \multicolumn{2}{c|}{\bfseries Documentation Available}                                                                                                                                                   \\ \hline
\multirow{2}{.2\columnwidth}{\bfseries Perception of importance} & Important and Documented (I\_D) \newline \textbf{Expected situation}                                           & Important and Not Documented (I\_ND)
\newline \textbf{An Indication of Technical} \\ \cline{2-3} 
                                          & Not Important and Documented (NI\_D)
                                          \newline \textbf{An Indication of Waste} & Not Important and Not Documented (NI\_ND)
                                          \newline \textbf{Expected Situation}         \\ \hline
\end{tabular}
\end{table}

\subsection{Research Questions} \label{Section: Research Questions}

As mentioned in section \ref{Section: Concepts and assumption}, we argue that if a quality attribute is perceived important, then it should be documented. We have, therefore, divided our analysis into the following research questions:

\textbf{RQ1:} \textit{Can we identify Technical Debt and Waste in requirements documentation (as interpreted in section \ref{Section: Concepts and assumption}) from the responses in the NaPiRE questionnaire?} This question expresses our overarching objective of understanding the juxtaposition between the perception of the importance of a quality attribute and if it has been documented. The question is framed in the Technical Debt metaphor, as it conveys our understanding that: ``If a quality attribute is considered important, then it should be documented". Any deviation in this direction should be interpreted as a project decision that, for whatever reason, lead the practitioners into not documenting a quality attribute they consider important (i.e., an expression of Technical Debt). Likewise, ``if a quality attribute is not considered important, then it need not be documented". Any deviation in this direction we consider as an indication of Waste, as the effort invested in documenting the quality attribute, could have been better spent elsewhere in the development lifecycle. \textbf{RQ1} is divided into:

\textbf{RQ1.1: }\textit{For which quality attributes do the practitioners' responses indicate Technical Debt?} Through this sub-question, we will explore practitioners' responses to the NaPiRE 2018 dataset an identify the quality attributes in which a deviation is present of a quality attribute is perceived important and yet, it is not documented (referred in the analysis as \textit{I\_ND}).

\textbf{RQ1.2: }\textit{For which quality attributes do the practitioners' responses indicate Waste?} Through this sub-question, we will explore the practitioners´ responses to the NaPiRE 2018 dataset and identify the quality attributes in which a deviation is present of a quality attribute that is not perceived as important and yet, it has been documented (referred in the analysis as \textit{NI\_D}).

\textbf{RQ2: }\textit{How does the practitioners' context influence the occurrence of Technical Debt and Waste in requirements documentation?} This second research question conveys our pre-conception that practitioners fail to document some quality attributes that they consider important. 
\textbf{RQ2} is divided into:

\textbf{RQ2.1: }\textit{How does the system type influence the occurrence of Technical Debt and Waste?} This question conveys our pre-conception that the type of system can have an influence on the perceived importance of a quality attribute, and therefore on the occurrence of Technical Debt or Waste. 

\textbf{RQ2.2: }\textit{How does the project size influence the occurrence of Technical Debt and Waste?} This question conveys our pre-conception that the size of the software project can have an influence on the perception of importance or the documentation needs of a quality attribute. 

\textbf{RQ2.3: }\textit{How does the type of development process influence the occurrence of Technical Debt and Waste?}. This question conveys our pre-conception that the development process type might have an influence on the perception of importance or the documentation needs of a quality attribute. 

\subsection{Data Extraction and Analysis Procedure} \label{sec:data-extraction-analysis}

We base our analysis on the NaPiRE 2018 dataset and, thus, have access to the corresponding raw data as well as the pre-processed codification of the questionnaire and answers. Table \ref{tab:items} presents the variables included in this research.

A total of 488 responses are recorded for the NaPiRE 2018 instance of the survey. All recorded responses are complete for variables v\_6 to v\_13 (perceived importance of quality attributes, see Table \ref{tab:items}) whereas only 455 responses are complete for variables v\_97 to v\_102, v\_303, v\_103 (documentation of requirements for quality attributes, see Table \ref{tab:items}). We removed 57 responses for incompleteness in other variables of interest. Therefore, the total number of responses considered for this research is 398. Table \ref{table:Answers by quality attribute} presents the distribution of responses in the aforementioned categories by the type of quality attribute. 

\begin{table}[h]
\caption{Distribution of responses by quality attribute}
\label{table:Answers by quality attribute}
\centering
\begin{tabular}{L{.18\columnwidth}R{.2\columnwidth}R{.2\columnwidth}R{.2\columnwidth}R{.2\columnwidth}}
\toprule
\textbf{Quality attribute} & \textbf{Technical Debt (I\_ND)} & \textbf{Waste (NI\_D)} & \textbf{I\_D} & \textbf{NI\_ND} \\
\midrule
Compatibility   &    70 (17.59\%) &     46 (11.56\%) &       99 (24.87\%) &    183 (45.98\%) \\
Maintainability &    123 (30.9\%) &      27 (6.78\%) &      105 (26.38\%) &    143 (35.93\%) \\
Performance     &    90 (22.61\%) &     47 (11.81\%) &      143 (35.93\%) &    118 (29.65\%) \\
Portability     &    46 (11.56\%) &       39 (9.8\%) &        31 (7.79\%) &    282 (70.85\%) \\
Reliability     &   122 (30.65\%) &      31 (7.79\%) &       117 (29.4\%) &    128 (32.16\%) \\
Safety          &    68 (17.09\%) &      31 (7.79\%) &         39 (9.8\%) &    260 (65.33\%) \\
Security        &     80 (20.1\%) &     59 (14.82\%) &      125 (31.41\%) &    134 (33.67\%) \\
Usability       &    97 (24.37\%) &      35 (8.79\%) &       158 (39.7\%) &    108 (27.14\%) \\ \midrule
Mean            &  87.0 (21.86\%) &  39.375 (9.89\%) &  102.125 (25.66\%) &  169.5 (42.59\%) \\
\bottomrule
\end{tabular}
\end{table}

The distribution of the contextual project information that will be analyzed for \textbf{RQ2} is shown in Table \ref{Table:Distribution of Responses}. It is worth mentioning that we applied a pre-processing step to variable v\_3 since it represents a free-text response. We used the results from the variable coding made by the collaborators of the NaPiRE initiative during their data analysis phase. For the purpose of analysing this variable, we grouped the responses into equal-sized buckets that represent small-sized (v\_3 $< 7$), medium-sized ($7 \leq$ v\_3 $< 15$) and large-sized projects (v\_3 $\geq 15$). 

\begin{table}[h]
\centering
\caption{Number of responses by variable under study. Mean and standard deviation are reported for the percentages of responses indicating Technical Debt and Waste, calculated across all the quality attributes.}
\label{Table:Distribution of Responses}
\begin{tabular}{L{.2\columnwidth}L{.25\columnwidth}R{.16\columnwidth}R{.18\columnwidth}R{.18\columnwidth}}
\toprule
\textbf{Variable} &  \textbf{Value} & \textbf{Responses} & \textbf{Technical Debt (I\_ND)} & \textbf{Waste (NI\_D)}  \\
\midrule
Process type & Agile &     $63$ &  $26.59 \pm 8.60$ &   $7.54 \pm 2.91$ \\
             & Hybrid &    $135$ &  $20.28 \pm 6.07$ &   $9.81 \pm 2.85$ \\
             & Plan-driven &     $37$ &  $22.30 \pm 9.88$ &   $8.78 \pm 4.51$ \\
             & Rather agile &     $95$ &  $23.16 \pm 9.87$ &  $10.26 \pm 4.96$ \\
             & Rather plan-driven &     $68$ &  $18.57 \pm 6.29$ &  $12.32 \pm 4.30$ \\
Project size & L &    $123$ &  $20.93 \pm 5.84$ &   $9.96 \pm 3.00$ \\
             & M &    $136$ &  $20.50 \pm 6.59$ &  $11.49 \pm 3.79$ \\
             & S &    $139$ &  $24.01 \pm 8.34$ &   $8.27 \pm 2.49$ \\
System class & BIS &    $202$ &  $22.40 \pm 8.87$ &   $9.65 \pm 4.23$ \\
             & HYB &    $101$ &  $23.14 \pm 5.50$ &   $8.91 \pm 2.80$ \\
             & SIES &     $95$ &  $19.34 \pm 5.39$ &  $11.45 \pm 2.42$ \\
\bottomrule
\end{tabular}
\end{table}


\section{Results} \label{sec:results}

This section shows the results of our analysis organized by the research questions.

\textbf{RQ1:} \textit{Can we identify Technical Debt and Waste (as interpreted in Section \ref{Section: Concepts and assumption}) from the responses in the NaPiRE questionnaire?} To answer RQ1, we cross-reference the responses to the perceived importance of quality attributes (v\_6 to v\_13) with the availability of documentation (v\_97 to v\_102, v\_303, and v\_103). Important and not documented (I\_ND) requirements indicate Technical Debt, whereas not important and documented (NI\_D) requirements indicate Waste (see Table~\ref{Table:Important_Documented}).

\textbf{RQ1.1: }\textit{For which quality attributes do the practitioners' responses indicate Technical Debt?} Table \ref{table:Answers by quality attribute} shows the occurrence of Technical Debt for each quality attribute. The percentage of responses showing Technical Debt (I\_ND) ranges from 12\% to 31\%. The average percentage of responses related to Technical Debt over all quality attribute types is 22\%. The quality attributes which are most likely to incur in Technical Debt are Reliability (31\%), Maintainability (31\%), Usability (24\%), and Performance (23\%).

\textbf{RQ1.2: }\textit{For which quality attributes do the practitioners' responses indicate Waste?} Table \ref{table:Answers by quality attribute} shows that waste also occurs in all quality attributes (albeit at a smaller response rate). The percentage of responses showing Waste (NI\_D) ranges from 7\% to 15\%. The  quality attributes which exhibit higher Waste are Security (15\%), Performance (12\%), and Compatibility (12\%).

\begin{figure*}[h]
\centering
\includegraphics[width=1\textwidth]{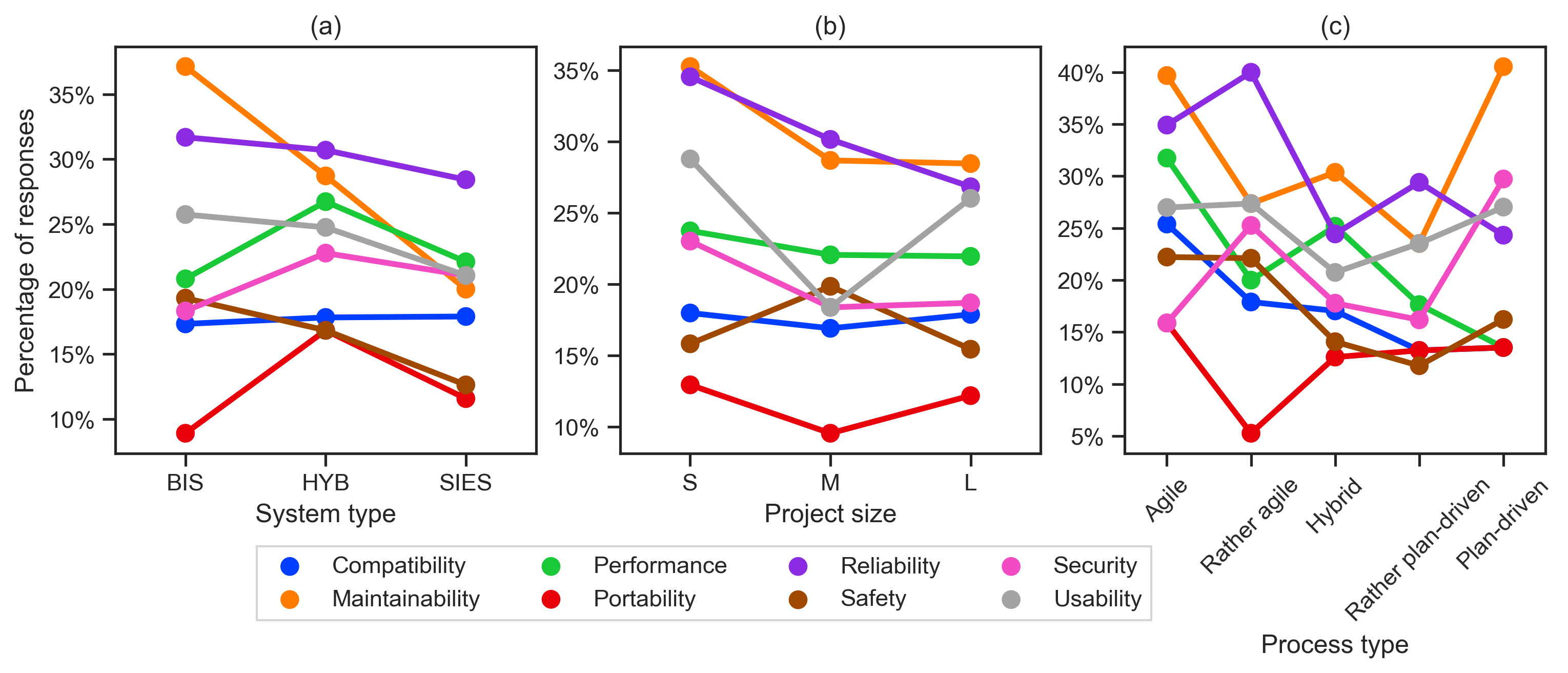}
\caption{Percentage of responses indicating Technical Debt by System type, Project size, and Process type.}
\label{Fig:TD by variable}
\end{figure*}

\begin{figure*}[h]
\centering
\includegraphics[width=1\textwidth]{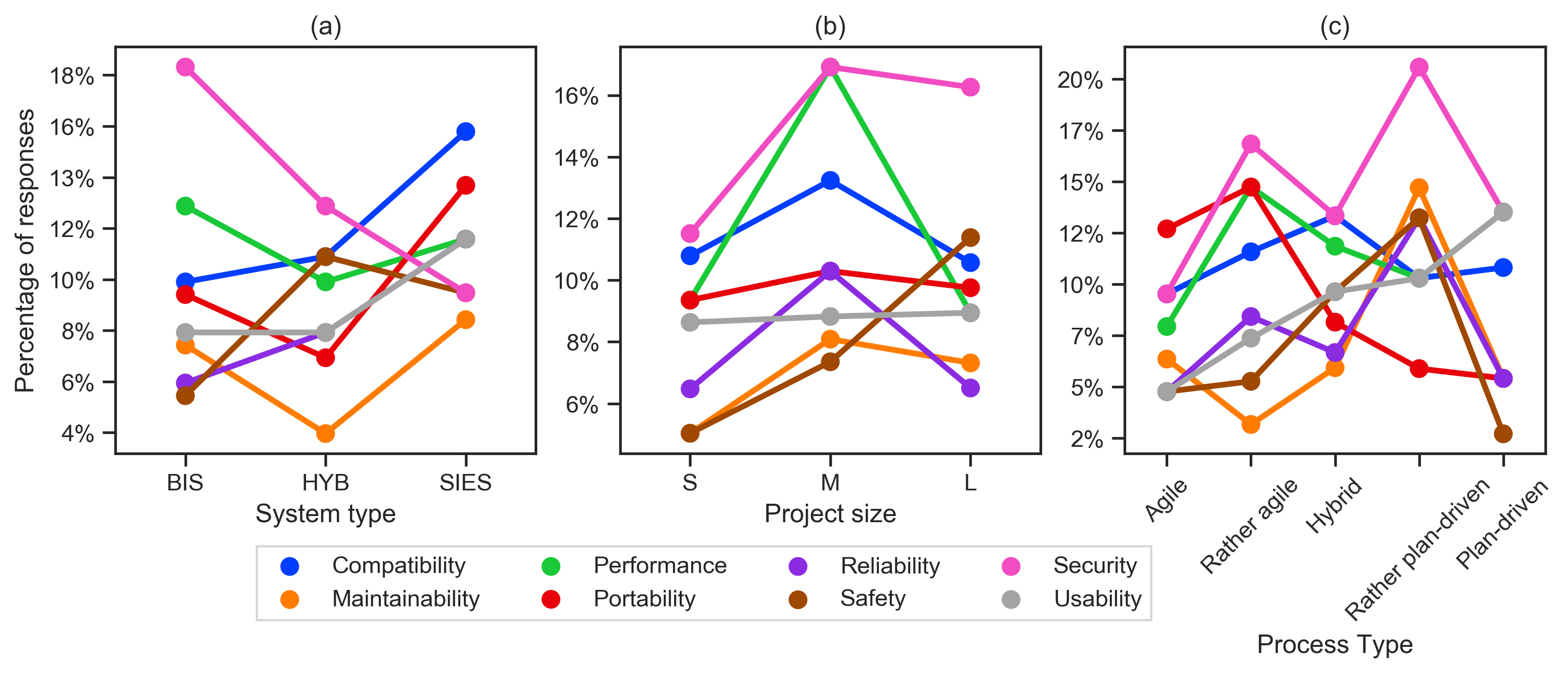}
\caption{Percentage of responses indicating Waste by System type, Project size, and Process type.}
\label{Fig:Waste by variable}
\end{figure*}

\textbf{RQ2: }\textit{How does the practitioners context influence the occurrence of Technical Debt and Waste?} To answer RQ2, we blocked the response data by the variables type of system, project size, and development process type to investigate their influence on the occurrence of Technical Debt and Waste. Figure \ref{Fig:TD by variable} shows the percentage of responses by quality attribute that indicate Technical Debt for each of the variables under study. Similarly, Figure \ref{Fig:Waste by variable} shows the percentage of responses related to Waste.

\textbf{RQ2.1: }\textit{How does the system type influence the occurrence of Technical Debt and Waste?}. When broken down by the system type (see Figure \ref{Fig:TD by variable} (a)) we can observe that Reliability is the most prone to Technical Debt in all types of systems. On the other end, Portability, is not prone to Technical Debt in the system types under analysis. The HYB type of system is type of system where the average percentage of responses indicating Technical Debt is the highest (23\%) (see Table \ref{Table:Distribution of Responses}). Four quality attributes surpass the average percentage of responses for all the quality attributes (22\%), namely Usability (25\%), Reliability (31\%), Performance (27\%), and Maintainability (29\%). Security (23\%) can be considered a borderline case. The percentage of BIS showing Technical Debt ranges from 9\% to 37\% with highest values for Maintainability (37\%), Reliability (32\%), and Usability (26\%). From Figure \ref{Fig:TD by variable} (a) we can see that BIS systems three quality attributes surpass the average percentage of responses for all the quality attributes (22\%), namely Usability (26\%), Reliability (32\%), and Maintainability (37\%). This system type also shows the highest percentages for Maintainability (37\%) and Reliability (32\%). Finally, SIES systems show percentages of Technical Debt ranging from 12\% to 28\%, and four quality attributes surpass the average percentage of responses (19\%), namely, Reliability (28\%), Performance (22\%), Security (21\%), and Usability (21\%). Maintainability (20\%) can be considered as a borderline case.

Regarding Waste, the average percentage of responses for all the quality attributes is 10\%, 9\%, and 11\% for BIS, HYB, and SIES type of systems (see Table \ref{Table:Distribution of Responses}). When broken down by the system type (see left-side of Figure \ref{Fig:Waste by variable} (a)), the highest percentages of responses are related to the Security of BISs (18\%) and the Compatibility of SIES (16\%). At the other end, the lowest percentage is related to the Maintainability of HYB systems (4\%). 

\textbf{RQ2.2: }\textit{How does project size influence the occurrence of Technical Debt and Waste?} Figure \ref{Fig:TD by variable} (b) shows the Technical Debt for each quality attribute blocked by project size (S, M, L). Similarly, Figure \ref{Fig:Waste by variable} (b) shows the Waste. The average percentages of responses indicating Technical Debt is 24\%, 20\%, and 21\% for projects of size S, M, and L, respectively (see Table \ref{Table:Distribution of Responses}). 

Maintainability, and Reliability are the quality attributes which show the highest percentages of Technical Debt (regardless of project size). On the other end, Portability is the quality attribute with the lowest Technical Debt regardless of project size. Small projects incur in Technical Debt having percentages ranging from 13\% to 35\%. This kind of projects particularly shows high percentages related to Reliability (35\%) and Maintainability (35\%). As for medium-sized projects, the percentages range from 10\% to 30\%. In large-sized projects, the percentages of Technical Debt range from 12\% to 28\%.

Regarding Waste, the average percentages of responses indicating Waste is 8\%, 11\%, and 10\% for projects of size S, M, and L, respectively (see Table \ref{Table:Distribution of Responses}). The percentages range from 5\% to 12\% for small-sized projects, from 7\% to 17\% for medium-sized projects, and from 7\% to 16\% for large-sized projects. In particular, the data points for Security and Safety seem to indicate that the number of responses showing Waste becomes larger as the project size increases. 

\textbf{RQ2.3: }\textit{How does type of development process influence the occurrence of Technical Debt and Waste?} Figure \ref{Fig:TD by variable} (c) shows the percentages of Technical Debt for every quality attribute organized by development process types. Similarly, Figure \ref{Fig:Waste by variable} (c) shows the percentages related to Waste. Three quality attributes exhibit the highest percentages of Technical Debt regardless of the type of development process, namely Maintainability, Reliability, and Usability. On the other hand, Portability is the only quality attribute without Technical Debt for any type of development process. 


Regarding Waste, the percentage responses indicating Waste related to Security is the highest. In addition, the percentage seems to increase as the projects become more plan-driven. The development process characterised as \textit{Rather plan-driven} shows the highest overall exposure to Waste.



\section{Discussion} \label{sec:discussion}

In this section we first discuss the results achieved regarding the relation between NFRs with Technical Debt and Waste, respectively. In addition, we discuss possible threats to validity of our study.\\

\textbf{Observations Related to NFR and Technical Debt}\label{Discussion:TD} Our results show that the majority of the participants of the survey stated that they document NFRs when they are important and they don't document NFRs when they are not important. This is what we had hypothesized. However, we observed that there is a substantial subset of respondents who stated that they don't document (some of the) important NFRs. This is what we interpret as being at risk of Technical Debt. Certain types of NFRs were particularly prone to this phenomenon, i.e., Reliability, Maintainability, Usability, and Performance. 
We can only speculate what would drive practitioners into this behaviour. For example, either these NFR types are difficult to document, knowledge on how to properly document NFRs is missing, or no appropriate tool is available. Furthermore, reasons might vary by NFR type. For example with \textit{Maintainability}, it can be argued that is left to good coding practices (i.e., avoiding code smells and focusing on refactoring). This phenomenon might also be true for other NFR types, i.e., there exist standard procedures or standard requirements that always hold and do not have to be explicitly stated in each individual project. When respondents answered the NaPiRE questionnaire, they might only have thought about project-specific documentation of NFRs.

\textbf{Observations Related to NFRs and Waste}\label{Discussion:TD} We also observed the occurrence of Waste, i.e., cases where respondents stated they document NFRs although they are not considered important. However, the observed Waste was consistently smaller than the Technical Debt (for the same quality attribute). Furthermore, when looking at the percentages observed for Waste, the proportion of Waste increases as project size increases: 5-11\% for small-sized projects, 7-17\% for medium-sized projects, and 7-16\% for large-sized projects. This might be a signal that - consistent with common expectation - for larger projects the risk of Waste is higher than for small projects with respect to NFRs. 
Surprisingly, and probably against common expectation, our analyses do not give any indication that projects using rather agile or purely agile processes produce less Waste than projects using plan-driven development approaches. 

\textbf{Threats to Validity}\label{sec:threats} We consider threats to construct, internal, external and conclusion validity according to Wohlin et al.~\cite{Wohlin2012} as well as measures to mitigate them.

This research is grounded on the NaPiRE 2018 survey, therefore we inherit some of the decisions taken during the development of the survey instrument. Of particular importance to the research presented in this paper is the fact that the NaPiRE 2018 survey does not differentiate between quality attribute and NFR. Both concepts are confounded in the questions on which we based our analysis. As a research team we have discussed this issue in depth and decided to accept this threat as it is in line with our shared understanding that (1) we cannot revert this decision; (2) we share the understanding that practitioners would probably not differentiate between both (and even for those who do, we can probably not guarantee a shared understanding). The latter argument is in line with the results of Eckardt et al. \cite{Eckhardt2016} (already mentioned in Section \ref{Section: Concepts and assumption}) that there is a large variety in the understanding of what is quality and what are NFRs. Continuing with inherited threats, external validity of our results highly depends on the profile of participants in the NaPiRE survey. The survey received overall 488 responses from all over the world and we have shown in a previous paper~\cite{wagner2019status} that there are no significant differences in the NaPiRE data with respect to different cultural regions. Furthermore, we analyzed the data also with respect to the system type, the project size and the type of development process. We therefore think that threats to external validity are low.

An important construct validity injected by the approach described in this work relates to how the metaphor of Technical Debt and the concept of Waste were introduced into the analysis of the data set. First of all, the NaPiRE survey makes no reference to these concepts. Secondly, there is a subtle but present gap between the formulation of the questions and our interpretation. It can be argued that "which quality attribute is of particular high importance?" (as asked in the survey) is not the same as asking "List all quality attributes that are important".  

Regarding internal validity, a limitation that we always have with survey research is that surveys can only reveal perceptions of the respondents that might not fully represent reality. However, the analysis stems from the well-validated NaPiRE questionnaire (see Section~\ref{subsec:naire}), which has continuously been improved based on piloting and the first two runs. Furthermore, we tried to be explicit in our decision about our data cleaning criteria (see Section~\ref{sec:data-extraction-analysis}) to be able to perform a thorough analysis.



\section{Conclusion} \label{sec:conclusion}
This paper explored the relationship between the level of importance and the degree of documentation for the NFR types Compatibility, Maintainability, Performance, Portability, Reliability, Safety, Security, and Usability. The analysis is based on the data collected during the most recent run of the NAPiRE survey. To analyze this relationship, we refer to the Technical Debt and Waste metaphors. To the best of our knowledge, this is the first publication in which these two concepts were explored in the context of NFRs. The starting point of our analysis was the assumption that important NFRs must be documented. If a project breaks this rule, then we interpret it as a possible source of Technical Debt. Likewise, we postulated that not important NFRs should not be documented. If a project breaks this rule, then we interpret it as a possible source of Waste.

Our analyses indicate that for four types of NFR (Maintainability, Reliability, Usability, and Performance) more than 22\% of the survey respondents who labelled the respective NFR type as important said that they did not document it. We interpret this as an indication that these NFR types have a higher risk of Technical Debt than other NFR types. Our analysis also indicates that the risk of Waste is less evident than the Risk of Technical Debt. Regarding Waste, NFR relating to Security exhibit the highest (about 15\%) number of respondents that say that they do not consider Security important, but do document related requirements. For the remaining NFR under analysis, the respondents indicate that the problem of Waste is much less evident (when compared to Technical Debt). Additional analyses indicate that our results are not sensitive to the type of system class, the project size, or the type of development process.

Overall, we conclude that, for specific NFR types (i.e., Maintainability, Reliability, Usability, and Performance), there is a clear indication that lack of documentation of important NFRs occurs regularly, pointing to the risk of Technical Debt. Regarding Waste, with the exception of Security, we conclude that the manifestation of Waste is not as clear as the manifestation of Technical Debt. We discussed several potential reasons for the occurrence of this phenomenon. However, investigating the true causes of Technical Debt and Waste requires more empirical research, which we consider as future work. 





\section*{Acknowledgments}\label{sec:acknowledgments}
The authors would like to thank all practitioners who took the time to respond to the NaPiRE survey as well as all colleagues involved in the NaPiRE project. The authors further acknowledge Dietmar Pfahl's contribution to research process described in this paper. Ezequiel Scott is supported by the Estonian Center of Excellence in ICT research (EXCITE), ERF project TK148 „IT Tippkeskus EXCITE“. Gabriela Robiolo is supported by Universidad Austral.

%
%
%
\bibliographystyle{splncs04}
\bibliography{References}

\end{document}